\documentclass[aps,pra,twocolumn,showpacs,floatfix,superscriptaddress]{revtex4-1}

\pdfoutput=1

\usepackage{amsmath}
\usepackage{amssymb}
\usepackage{graphicx}
\usepackage{epstopdf}
\usepackage{latexsym}

\begin{document}

\title{Two-orbital physics of high spin fermionic alkaline-earth atoms confined in a one-dimensional chain}
\author{E. Szirmai}
\affiliation{BME-MTA Exotic Quantum Phases Research Group, Institute of Physics, Budapest University of Technology and Economics, 
Budafoki \'ut 8., H-1111 Budapest, Hungary}

\begin{abstract}
We study the effect of the coupling between the electronic ground state of high spin alkaline-earth fermionic atoms and their metastable optically excited state, when the system is confined in a one-dimensional chain, and show that the system provides a possible realization of a finite momentum pairing (Fulde-Ferrell-Larkin-Ovchinnikov-like) state without spin- or bare mass imbalance. We determine the $\beta$-functions of the renormalization group trajectories for general spin and analyze the structure of the possible gapped and gapless states in the hydrodynamic limit. Due to the SU(N) symmetry in the spin space, complete mode separation can not be observed even in the fully gapless 2N-component Luttinger liquid state. Contrary, 4 velocities characterize the system. We solve the renormalization group equations for spin-9/2 strontium-87 isotope and analyze in detail its phase diagram. The fully gapless Luttinger liquid state does not stabilize in the two-orbital system of the $^{87}$Sr atoms, instead, different gapped non-Gaussian fixed points are identified either with dominant density or superconducting fluctuations. The superconducting states are stable in a nontrivial shaped region in the parameter space as a consequence of the coupling between the two electronic states. 
\end{abstract}

\date{\today}
\maketitle

\section{Introduction}

Many fundamental questions and phenomena of quantum physics, magnetism or strongly correlated systems can be understood by studying the properties of spin systems. In the last decade ultracold atom experiments showed rapid progress, and today they provide a realistic possibility to study the consequences of high spin \cite{lewenstein07a,bloch08a}. Accordingly, the interest of high spin systems started to increase rapidly giving a new impulse to their theoretical investigations. Additionally, experiments with alkaline-earth atoms allow to study systems with very high symmetry: between alkaline-earth atoms the scattering processes have an SU(N) symmetry (N~$=2F+1$, and $F$ is the hyperfine spin of the atoms) within a very good accuracy due to the decoupling of the nuclear spin from the total electronic angular momentum. 
In the Mott regime, when the interaction is strongly repulsive, these systems can be described by an effective SU(N) spin-exchange model. These models --- especially on two dimensional lattices ---  can provide a series of nontrivial states depending on the value of N and the geometry of the underlying lattice. Like different bond- and site-centered magnetic orders, valence bond, plaquette or spin liquid states, or even chiral spin liquid states with nontrivial topology \cite{Read1989,Marston1989,harada03a,Assaad2005,chen05a,tu06a,Paramekanti2007,Greiter2007,xu08a,hermele09a,wu10b,toth10a,corboz11a,szirmai11a,szirmai11b,hung11a,yang12a,corboz12a,lajko13a}.  In the attractive regime special superfluid states can emerge as a direct consequence of high spin, like multiparticle (trion, quartet, etc.) superfluidity or mixed superfluid phases in which Cooper-like pairs carrying different magnetic moments coexist \cite{ho99a,yip99a,wu03a, rapp07a,wu10c,barcza12a}.

One-dimensional high spin systems have also been studied intensively \cite{szirmai05a,capponi07a,buchta07a,szirmai08a,lecheminant08a, capponi08a,azaria09a, nonne10a,manmana11a,nonne11a, nonne11b,duivenvoorden13a,duivenvoorden13b}, basically within the framework of the generalization of the Hubbard and Heisenberg models. The special case of spin-3/2 fermions as the simplest one beyond the usual spin-1/2 electron system has been studied extensively, and now we have a rather detailed knowledge of this system  \cite{wu03a,wu05a,wu06a,capponi07a,guan09a, barcza12a,nonne13a}. With the help of bosonization, and analytical renormalization group (RG), one can characterize some special features of the high spin systems for general N, too. For instance, it was shown that in the SU(N) Hubbard chain at incommensurate fillings a generalization of the spin-charge separation, namely, total mode separation occurs, and the system is equivalent to an N-component Luttinger liquid. Contrary, at half filling even the usual spin-charge separation breaks down, if N~$>2$ \cite{szirmai05a,buchta07a}. The further details of the realized phase always depend on the value of N, therefore, for a specific N, further investigation is needed to identify the emerging states.

An additional internal degree of freedom, like orbital state or internal electronic state of atoms, can essentially change the properties of the realized phase. For spin-1/2 electron systems the effect of an additional two-state internal degree of freedom was studied widely, usually by forming the problem suitable to describe specific condensed matter systems. In these works the coupling between the two orbital states occurs as weak hybridization effect \cite{varma85a,penc90a} that makes the problem analogous with a two-leg Hubbard ladder problem even in the high spin case \cite{szirmai06a}. 
Despite, currently we have a quite poor knowledge about the two-orbital physics of high spin fermions. 
In ultracold atomic systems, instead of hybridization, the most important coupling effects come from the scattering processes between particles in different orbital states. In Ref. \cite{gorshkov10a} the authors proposed a fundamental model for the description of the related ultracold atomic experiments. They also gave a detailed description of two-orbital SU(N) magnetism on two dimensional lattices, in the partly localized and in the Mott state. They pointed out that in the strong repulsive case the two-orbital model can be used to implement important models of condensed matter systems, like Kugel-Khomskii model used to describe spin-orbital physics in transition metal oxides \cite{Kugel1973,*Arovas1995,*Li1998,*Pati1998,*Tokura2000}, or the Kondo lattice model often used to describe heavy fermion materials \cite{Ruderman1954, *Coqblin1969, *Doniach1977,*Coleman1983,*Tsunetsugu1997,*Assaad1999,*Tokura2000,*Oshikawa2000,*Senthil2003,*Duan2004,*Paredes2005,*Coleman2007,*Gegenwart2008}. 
In Ref.~\cite{nonne13a} the half-filled two-orbital SU(N) chain has been analyzed from aspects of the possible topological phases in the Mott regime.
As one goes farther on from the localized states, due to the strong competition of the kinetic energy and the potential energy, the Mott state is melted, and the emerging states are difficult to describe. In Ref.~\cite{xu10a} C. Xu analyzed the $k$-orbital system in a quite general way and gave a classification of the quantum liquid states based on the coupling of the orbital, spin and charge fluctuations.

In this paper we study the two-orbital physics of one dimensional SU(N) fermionic atoms far from the Mott state. The orbital degree of freedom is mimicked by two metastable electronic states \cite{gorshkov10a}.  After discussing general features of a one-dimensional chain of two-orbital atoms, we present the phase diagram of the $^{87}$Sr isotopes. We find that the system of $^{87}$Sr atoms can not show a fully gapless Luttinger liquid behavior, only gapped states can stabilize, even at incommensurate fillings. The phase boundaries between the density wave phase and the superconducting phase have complex structure as a consequence of the competition between the various dressed interorbital interactions. Due to the effective mass imbalance between the ground state and the optically excited state, the Cooper pairs in the superconducting phase have a finite momentum, similar to the Fulde-Ferrell-Larkin-Ovchinnikov (FFLO) state \cite{fulde64a,*larkin64a}. 

The structure of the paper is the following: In Sec.~\ref{sec:model} the model is presented and the notations are introduced. In Sec.~\ref{sec:general-properties} a general analysis is given in the hydrodynamic regime, where the bosonization treatment is reliable. We determine the RG equations for general N, and analyze the general properties of the Luttinger liquid phase and the gapped phases in boson representation. In Sec.~\ref{sec:Sr} we solve numerically the RG equations for the special case of the $^{87}$Sr isotope to determine its phase diagram, than in Sec.~\ref{sec:experiment} we discuss some aspects of the experimental study the system. In the last section we give a short summary and conclusion of the results.

\section{Formulation of the problem}
\label{sec:model}

In what follows we consider a fermionic system with hyperfine spin $F$ loaded into a one-dimensional optical lattice. The atoms can be driven from their electronic ground state ($^1S_0$) $|g\big>$ to a metastable exited state ($^3P_0$) $|e\big>$ as it was introduced in Ref. \cite{gorshkov10a}. Note, that the total electronic angular momentum remains 0 in the excited state, too.
Accordingly, the non-interacting terms of the Hamiltonian of the effective two-orbital system reads as $H_0=H_0^g + H_0^e + H_0^{ge}$, where the intraorbital tunneling is
\begin{equation}
\label{eq:ham-kin}
H_{0}^{\alpha}=-\sum_{i,\sigma} t_\alpha ( c_{i,\alpha,\sigma}^\dagger c_{i+1,\alpha,\sigma} + H.c.) , 
\end{equation}
with $\alpha=g$ or $e$. Here and in the following $c_{i,\alpha,\sigma}^\dagger$ ($c_{i,\alpha,\sigma}$) creates (annihilates) an atom in the orbital state $\alpha$ with spin $\sigma$ on site $i$. 
The hopping amplitudes within a tight-binding approximation is $t_\alpha = - \int \mathrm{d} \mathbf{r} w_\alpha^*(\mathbf{r})\Big(-\frac{\hbar^2}{2M_\textrm{atom}} \nabla + V_\alpha (\mathbf{r})\Big)w_\alpha(\mathbf{r} - a \mathbf{e})$, where 
$w_\alpha (\mathbf{r})$ is the Wannier function of the particles, $\mathbf{e}$ denotes the unit vector along the chain, and $a$ is the lattice constant of the underlying optical lattice.
$V_\alpha(\mathbf{r})$ describes the optical lattice potential with one-dimensional periodicity: $V_\alpha(\mathbf{r})=V_\alpha(\mathbf{r} + m a \mathbf{e})$ with arbitrary integer $m$. Generally, it shows a weak parabolic site dependence that is neglected in the following, and we assume that the lattice potential does not couple to the nuclear spin.  

While the typical values of the intraorbital tunneling is in the order of Hz-kHz, the transition frequency between the ground state and the excited state is in the optical range ($\sim$ 1000 THz). Therefore, the transition between the two states is off-resonant, the hybridization of the metastable excited state with the ground state can be neglected \cite{gorshkov10a}. The lifetime in the metastable state is relatively long, usually a few ms, therefore, without interaction or in case of elastic scatterings the population of the ground state and of the excited state can be considered as fixed.  
The energy difference according to the population of the excited state can be described as:
\begin{equation}
\label{eq:ham-electr.excitation}
H_0^{ge}=\hbar \omega_0 \sum_{i,\sigma} \left[ n_{i,e,\sigma} - n_{i,g,\sigma} \right].
\end{equation}
Here $n_{i\alpha,\sigma}$ is the particle number operator: $n_{i,\alpha,\sigma}=c_{i,\alpha,\sigma}^\dagger c_{i,\alpha,\sigma}$. This term gives a constant shift to the energy proportional to the excitation energy $\hbar \omega_0$ at fixed occupation of the two orbital states. 

The fermions interact decisively via a weak Van der Waals interaction that can be approximated with an effective $s$-wave contact potential. The $s$-wave scattering length depends on the electronic states of the colliding atoms, but it is independent of the hyperfine spin in case of alkaline-earth atoms. This latter property is a consequence of the closed electronic shell structure in which case the total electronic angular momentum of the atom is zero. Therefore the hyperfine spin comes only from the nuclear spin that does not affect the Van der Waals interaction. This leads to an SU(N) symmetry of the interaction in the spin space. Accordingly, four independent couplings characterize the atomic interaction: $g_g$ ($g_e$) when both colliding particles are in the ground state (excited state), and $g_{ge}^+$ ($g_{ge}^-$) when one of the scattering particles is in the ground state, the other is in the excited state, and the two-particle state is symmetric (antisymmetric) in the electronic state. The couplings can be tuned via the corresponding $s$-wave scattering length $a_{g(e)}$, and $a_{ge}^\pm$ as $g_{g(e)} \approx 4 \pi \hbar^2 a_{g(e)} \mathcal{I}_{g(e)}/M_\textrm{atom}$, and $g_{ge}^{\pm} \approx 4 \pi \hbar^2 a_{ge}^{\pm}\mathcal{I}_{ge} /M_\textrm{atom}$, respectively. The interaction also depends on the parameters of the underlying lattice via the integrals $\mathcal{I}_{g(e)}=\int \mathrm{d}\mathbf{r} [w_{g(e)}^*(\mathbf{r})w_{g(e)}(\mathbf{r})]^2$, and $\mathcal{I}_{ge} = \int \mathrm{d}\mathbf{r} w_g^*(\mathbf{r}) w_g(\mathbf{r})w_e^*(\mathbf{r}) w_e(\mathbf{r})$.
Accordingly, the intraorbital scatterings can be described by simple density-density interaction:
\begin{equation}
\label{eq:ham-int-m}
H_\mathrm{int}^\alpha =  \frac12  g_\alpha  \sum_i \sum_{\sigma \neq \sigma'} n_{i,\alpha,\sigma}n_{i,\alpha,\sigma'},
\end{equation}
and the coupling between the electronic states $|g\big>$ and $|e\big>$ contains density-density interaction and exchange term:
\begin{multline}
\label{eq:ham-int-ge}
H_\mathrm{int}^{ge} = \frac12 \sum_{i,\sigma,\sigma'} \Big[ g_{ge}\, n_{i,e,\sigma}n_{i,g,\sigma'} \\ + g_{ge}^{\textrm{ex}} \,
 c_{i,g,\sigma}^\dagger c_{i,e,\sigma'}^\dagger c_{i,g,\sigma'} c_{i,e,\sigma} \big].
\end{multline}
Here $g_{ge}=g_{ge}^+ + g_{ge}^-$, and $g_{ge}^{\textrm{ex}}=g_{ge}^+ - g_{ge}^-$. 

Since the interaction strength does not depend on the hyperfine spin state of the scattering particles, the density-density interaction terms just as $H_0^{ge}$ in Eq. \eqref{eq:ham-electr.excitation}, have local SU(N) symmetry, independently on each other in the two electronic states. 
This local symmetry shows that the particle number and the SU(N) spin in both electronic states and on each site are preserved by these terms. The locality of this symmetry is violated by the hopping terms, therefore, without exchange interaction the system has the global SU$_g$(N)$\times$SU$_e$(N) symmetry, corresponding to the SU(N) spin rotational invariance, independently in the electronic ground state and the excited state. 
The exchange interaction between two particles with different spin states does not preserve the independent SU(N) invariance in the two electronic state. It couples the spins in the $|g\big>$ and $|e\big>$ sates and violates the SU$_g$(N)$\times$SU$_e$(N) symmetry to SU(N).

\section{Continuum limit}
\label{sec:general-properties}

The low energy physics of the system can be well described within hydrodynamical approach. Therefore, first we construct the corresponding continuum model. The population of the two electronic states determines the Fermi surface that consists four Fermi points $\pm k_\mathrm{F}^g$ and $\pm k_\mathrm{F}^e$ in the one-dimensional case. Around these Fermi-points the spectrum can be linearized leading to four well separated branches of the low energy spectrum. Introducing the corresponding operators $L_{\alpha,\sigma}(x)$ and $R_{\alpha,\sigma}(x)$ of the left and right moving particles ($x$ denotes the continuous space coordinate along the chain), the continuum limit can be done by the exchange 
\begin{equation}
\label{eq.cont-lim}
\frac{1}{\sqrt{a}} c_{i,\alpha,\sigma}\rightarrow L_{\alpha,\sigma}(x) \mathrm{e}^{-\mathrm{i}k_\mathrm{F}^\alpha x} +  R_{\alpha,\sigma}(x) \mathrm{e}^{\mathrm{i}k_\mathrm{F}^\alpha x}.
\end{equation}
Now the kinetic term can be written into the following form:
\begin{equation}
\label{eq:ham-kin-fermionrep}
H_0=-\mathrm{i} \sum_{\alpha,\sigma} \int \mathrm{d} x v_\alpha (R_{\alpha,\sigma}^\dagger \partial_x R_{\alpha,\sigma} - L_{\alpha,\sigma}^\dagger \partial_x L_{\alpha,\sigma}),
\end{equation}
where $v_\alpha=2a t_\alpha \sin{(k_\mathrm{F}^\alpha a)}$. Since we work with fixed number of particles in the two excited states, the term \eqref{eq:ham-electr.excitation} gives only an uninteresting constant to the energy.

\begin{figure}
\includegraphics[scale=0.3]{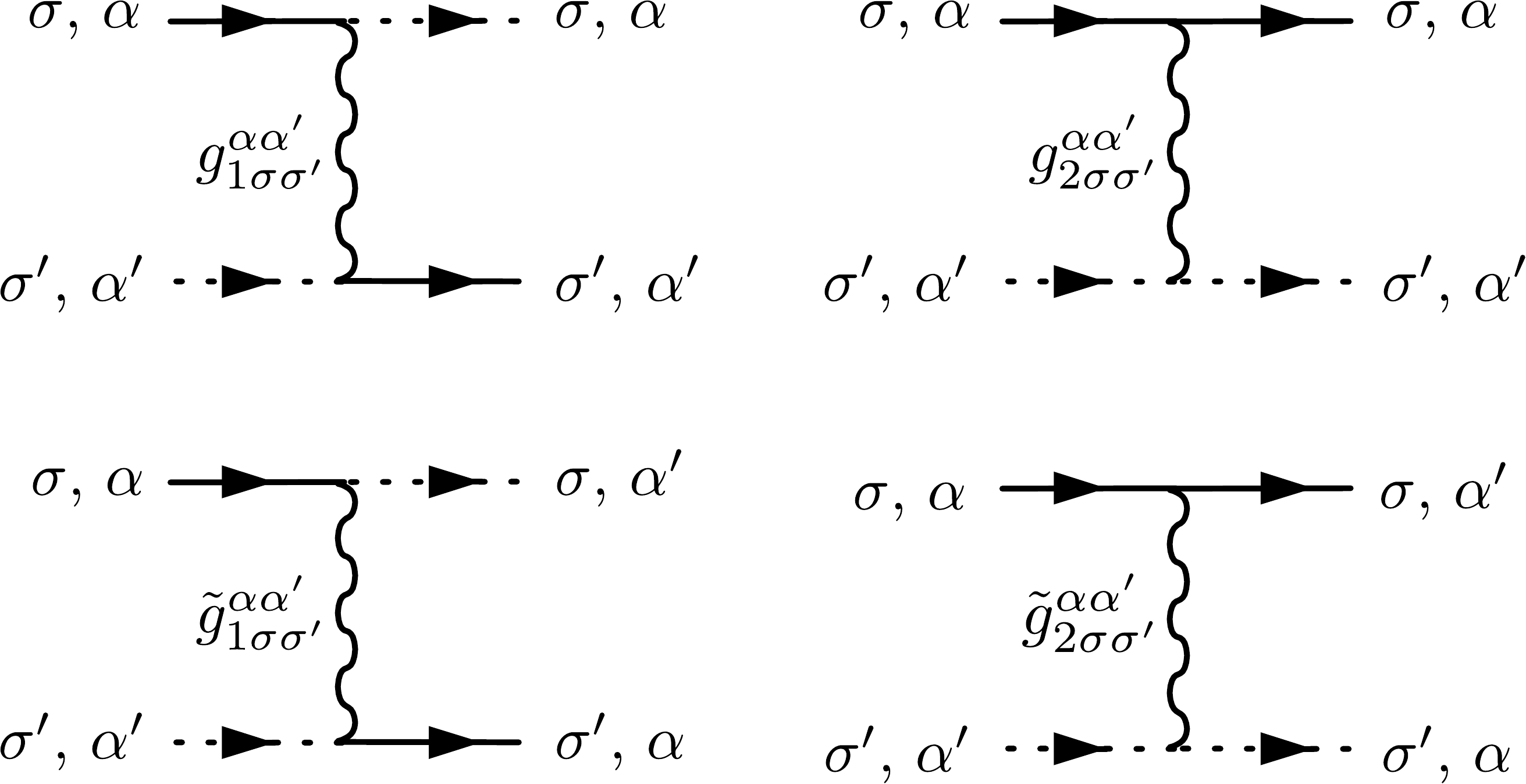}
\caption{The various bare vertices corresponding to the definition in Eq.~\eqref{eq:gamma-def}. The solid lines refer to right moving particles, the dashed lines refer to left moving particles, and the wavy lines denote the interaction. With these definitions none of the interactions flips the spin, while the $\tilde{g}$ scatterings exchange the orbital states.}
\label{fig:bare-vertices}
\end{figure}

The scattering processes can be classified by the momentum transfer between the colliding particles, and by the change of their internal (spin and electronic) state. Away from half filling, the two particle umklapp processes are irrelevant, therefore the only interesting processes take place between a left and a right moving particles. Let us label their internal states with the subscripts $il$, $ir$, $fl$, and $fr$, as initial left, initial right moving, and final left, final right moving particles. Considering that none of the interaction terms flips the spin state, but $H_\mathrm{int}^{ge}$ exchange the electronic state of the two scattering particles one can define the following scattering vertices:
\begin{subequations}
\label{eq:gamma-def}
\begin{flalign}
\Gamma_1 (\{\alpha, \sigma \})\, \delta_{\alpha_{ir},\alpha_{fl}} \delta_{\sigma_{ir},\sigma_{fl}} \delta_{\alpha_{il},\alpha_{fr}}  \delta_{\sigma_{il},\sigma_{fr}}, \\
\Gamma_2 (\{\alpha, \sigma \}) \, \delta_{\alpha_{ir},\alpha_{fr}} \delta_{\sigma_{ir},\sigma_{fr}} \delta_{\alpha_{il},\alpha_{fl}} \delta_{\sigma_{il},\sigma_{fl}}, \\
\tilde{\Gamma}_1 (\{\alpha, \sigma \}) \, \delta_{\alpha_{ir},\alpha_{fr}} \delta_{\sigma_{ir},\sigma_{fl}} \delta_{\alpha_{il},\alpha_{fl}} \delta_{\sigma_{il},\sigma_{fr}}, \\
\tilde{\Gamma}_2 (\{\alpha, \sigma \}) \, \delta_{\alpha_{ir},\alpha_{fl}} \delta_{\sigma_{ir},\sigma_{fr}} \delta_{\alpha_{il},\alpha_{fr}}  \delta_{\sigma_{il},\sigma_{fl}}.
\end{flalign}
\end{subequations}
The vertices $\Gamma_1$ and $\tilde{\Gamma}_1$ describe scatterings with momentum transfer $\pm(k_\mathrm{F}^{\alpha_{ir}} + k_\mathrm{F}^{\alpha_{il}})$, while during the $\Gamma_2$ and $\tilde{\Gamma}_2$-type processes the momentum transfer is $\pm(k_\mathrm{F}^{\alpha_{ir}} - k_\mathrm{F}^{\alpha_{il}})$. With these definition the vertices are well-defined, and due to the Kronecker deltas they can be characterized by simply the spin and orbital parameters of the incoming right ($\alpha_{ir}$, $\sigma_{ir}$) and left 
($\alpha_{il}$, $\sigma_{il}$) moving particles. The corresponding bare interaction vertices are denoted by $g_{1\sigma_{il}\sigma_{ir}}^{\alpha_{il} \alpha_{ir}}$, $g_{2\sigma_{il}\sigma_{ir}}^{\alpha_{il} \alpha_{ir}}$, $\tilde{g}_{1\sigma_{il}\sigma_{ir}}^{\alpha_{il} \alpha_{ir}}$, and $\tilde{g}_{2\sigma_{il}\sigma_{ir}}^{\alpha_{il} \alpha_{ir}}$, and in Fig.~\ref{fig:bare-vertices} their diagrammatic notation are shown. From the above definitions it is obvious that the $\tilde{\Gamma}$ processes between atoms with either parallel spin or in the same orbital state do not determine new processes, therefore we do not define these processes.

To study the relevance of the various scattering processes we used renormalization group treatment. Applying the RG procedure  \cite{solyom79a} for all the interaction terms listed above, one can arrive to the $\beta$-functions of a very general system, namely, for general spin and orbital dependence of the couplings in Eqs.~\eqref{eq:ham-int-m}-\eqref{eq:ham-int-ge}. 
The $\beta$-functions of the renormalization group procedure can be determined based on perturbation theory. Up to the leading one-loop order they have the following form (for the details of the calculations see Appendix A):
\begin{subequations}
\label{eq:betafunctions}
\begin{flalign}
\nonumber 
\beta_{1\sigma \sigma'}^{\alpha \alpha'}  = & \,\,
2 \, \big( g_{1\sigma \sigma'}^{\alpha \alpha'} g_{2\sigma \sigma'}^{\alpha \alpha'} + \tilde{g}_{1\sigma \sigma'}^{\alpha \alpha'} \tilde{g}_{2\sigma \sigma'}^{\alpha \alpha'} \big)/\pi(v_{\alpha} + v_{\alpha'}) \\ 
\nonumber
 & - \big( g_{2\sigma \sigma}^{\alpha \alpha} g_{1\sigma \sigma'}^{\alpha \alpha'} + g_{1\sigma \sigma'}^{\alpha \alpha} \tilde{g}_{1\sigma' \sigma'}^{\alpha \alpha'} \big)/2\pi v_\alpha  \\ 
 \nonumber
& - \big( \tilde{g}_{1\sigma \sigma}^{\alpha \alpha'} g_{1\sigma \sigma'}^{\alpha' \alpha'} + g_{1\sigma \sigma'}^{\alpha \alpha'} g_{2\sigma' \sigma'}^{\alpha' \alpha'} \big)/2\pi v_{\alpha'} \\ 
\label{eq:beta-a}  
& + \sum_{\hat{\sigma},\hat{\alpha}} 
 g_{1\sigma \hat{\sigma}}^{\alpha \hat{\alpha}} g_{1\hat{\sigma} \sigma'}^{\hat{\alpha} \alpha'}/2\pi v_{\hat{\alpha}} ,\\
 \beta_{2\sigma \sigma'}^{\alpha \alpha'}  = & \,\, \big( g_{1\sigma \sigma'}^{\alpha \alpha'} g_{1\sigma \sigma'}^{\alpha \alpha'} + \tilde{g}_{2\sigma \sigma'}^{\alpha \alpha'} \tilde{g}_{2\sigma \sigma'}^{\alpha \alpha'} \big) / \pi(v_{\alpha} + v_{\alpha'})  ,   \\
 \tilde{\beta}_{1\sigma \sigma'}^{\alpha \alpha'}  = & \,\, 2\, g_{1\sigma \sigma'}^{\alpha \alpha'}  \tilde{g}_{2\sigma \sigma'}^{\alpha \alpha'}/\pi(v_{\alpha} + v_{\alpha'}), \\
\nonumber 
\tilde{\beta}_{2\sigma \sigma'}^{\alpha \alpha'}  = & \,\,\big(  2\, g_{1\sigma \sigma'}^{\alpha \alpha'} \tilde{g}_{1\sigma \sigma'}^{\alpha \alpha'} + 2 \, \tilde{g}_{2\sigma \sigma'}^{\alpha \alpha'}  g_{2\sigma \sigma'}^{\alpha \alpha'} \\ 
& -  \tilde{g}_{2\sigma \sigma'}^{\alpha \alpha'}  g_{2\sigma \sigma}^{\alpha \alpha'} -  \tilde{g}_{2\sigma \sigma'}^{\alpha \alpha'}  g_{2\sigma' \sigma'}^{\alpha \alpha'} \big)/\pi(v_{\alpha} + v_{\alpha'}).
\end{flalign}
\end{subequations}

Up to now we only assumed that the interactions do not flip the spins but they can depend on the spin of both scattering particles.
In the following, due to the SU(N) symmetry of the Hamiltonian, it is unnecessary  to keep the explicit spin dependence of the vertices and couplings, only its relative value is important. Therefore, we introduce the notation $\parallel$ and $\perp$, respectively, as subscript for the spin dependence of the different quantities. Similarly, the vertices are invariant under the exchange of their two orbital indices, therefore, the processes can be classify into three different channel considering the orbital state of the scattering particles: either both atoms are in the ground state (superscript $g$), or both are in the excited state (superscript $e$), or one is in the ground state and the other is in the excited state (superscript $ge$). It is worth to emphasize that in the SU(N) symmetric case the N dependence of the $\beta$-function occurs only in Eq.~\eqref{eq:beta-a} because of the summation over $\hat{\sigma}$. 
The initial values of the couplings in the two-orbital system described by Hamiltonian in Eqs.~\eqref{eq:ham-kin}-\eqref{eq:ham-int-ge} are 
\begin{subequations}
\label{eq:RG-initial_values}
\begin{flalign}
& g_{1\parallel}^e(0)=g_{1\parallel}^g(0)=g_{2\parallel}^e(0)=g_{2\parallel}^g(0)=0, \\ 
& g_{1\parallel}^{ge}(0)=g_{2\parallel}^{ge}(0)=g_{ge}^- , \\
& g_{1\perp}^{ge}(0)=g_{2\perp}^{ge}(0)=g_{ge}^+ + g_{ge}^- , \\
& g_{1\perp}^e(0)=g_{2\perp}^e(0)=g_e, \\ 
& g_{1\perp}^g(0)=g_{2\perp}^g(0)=g_g, \\
& \tilde{g}_{1\parallel}^{ge}(0)=\tilde{g}_{2\parallel}^{ge}(0)=0, \\ 
& \tilde{g}_{1\perp}^{ge}(0)=\tilde{g}_{2\perp}^{ge}(0)=g_{ge}^+ - g_{ge}^- .
\end{flalign}
\end{subequations}

Unfortunately, currently, rather few experimental data are available for the various scattering lengths, especially for the electronically excited states, and the complete analysis of the four-dimensional parameter space is actually out of feasibility. Nevertheless, as soon as any experimental data becomes available, with Eqs. \eqref{eq:betafunctions} and \eqref{eq:RG-initial_values} it is straightforward to study the fixed point structure and scaling trajectories providing a basis for further analysis of the possible phases. As a demonstration, in the next Section we apply our results to a specific isotope, the $^{87}$Sr, in an experimentally accessible regime. 

With the analysis of the RG equations one can determine the relevant scattering processes, but that does not provide information about their specific role. Within bosonization treatment, it is easy to classify these processes based on how they couple the various modes. 
Therefore, we will use the bosonized version of the Hamiltonian in Eqs.~\eqref{eq:ham-kin}-\eqref{eq:ham-int-ge} to describe some general properties of the two-orbital high spin fermionic system. In the field theoretical description \cite{gogolin04a,vondelft98a,giamarchi04a} one can use the following identity to define the boson fields and their canonically conjugated momentum fields:
\begin{subequations}
\label{eq:boson_fields}
\begin{flalign}
R_{\alpha,\sigma}(x) & = \frac{1}{\sqrt{2\pi a}} \mathrm{K}_{\alpha,\sigma} \mathrm{e}^{\mathrm{i}(\phi_{\alpha,\sigma}(x) + \theta_{\alpha,\sigma}(x))}, \\ 
L_{\alpha,\sigma}(x) & = \frac{1}{\sqrt{2\pi a}} \mathrm{K}_{\alpha,\sigma} \mathrm{e}^{-\mathrm{i}(\phi_{\alpha,\sigma}(x) - \theta_{\alpha,\sigma}(x))} .
\end{flalign}
\end{subequations}
Here $\mathrm{K}_{\alpha,\sigma}$ are the Klein factors to ensure the anticommutation relations of the fermionic fields $L_{\alpha,\sigma}$, and $R_{\alpha,\sigma}$, and $\theta_{\alpha,\sigma}$ are the dual fields of the bosonic phase fields $\phi_{\alpha,\sigma}$. The dual fields define the $\Pi_{\alpha,\sigma}$ canonical momentums conjugated to $\phi_{\alpha,\sigma}$ as $\Pi_{\alpha,\sigma}(x) = -\partial_x\theta_{\alpha,\sigma} (x)/\pi$.

The phases of a one-dimensional fermion system can be characterized by these bosonic fields. In general, some of them are pinned by the relevant interactions and can be excited only with a finite energy, while the others can fluctuate freely. The low energy excitations are always determined by the free (gapless) modes. At the same time, the emerging phases are characterized by the gapped modes, too. To illustrate this one can consider the half-filled SU(2) fermionic Hubbard chain, whose ground state is a spin liquid state (gapless spin mode) above a Mott insulating state (gapped charge mode). In the following we analyze how the relevance of the different interaction processes affect the behavior of the various modes --- separately for the fully gapless Luttinger liquid state and the various gapped states.

\subsection{Luttinger liquid state}
\label{sec:LL}

In the Luttinger liquid state all the 2N bosonic fields can fluctuate freely, their excitation spectrum is sound-like in the long wavelength limit, i.e. gapless and linear. 
In this state only the scattering processes $g_{1\parallel}^e$, $g_{1\parallel}^g$, $g_{2\parallel(\perp)}^e$, $g_{2\parallel(\perp)}^g$, and $g_{2\parallel(\perp)}^{ge}$, which preserve the spin and particle number at each branch of the spectra, can be relevant.  Accordingly, the system has the considerably high SU$_{L,g}$(N)$\times$SU$_{L,e}$(N)$\times$SU$_{R,g}$(N)$\times$SU$_{R,e}$(N) symmetry in the 2N component Luttinger liquid state. This is a Gaussian fixed point in which the Hamiltonian is quadratic and its diagonalization in the spin space can be performed with the help of the N$-1$ Cartan generators of the SU(N) and the N dimensional identity matrix. The definition of the Cartan generators can be found in Appendix~\ref{sec:app-a}. Note, that the spin-symmetric combination of the fields defined in Eq.~\eqref{eq:Cartan-0} usually referred as charge mode, because of its analogue in the electron system, and similarly, the combinations defined by the Cartan generators in Eq.~\eqref{eq:Cartan-l}, often called spin, or spin-like modes. In the following we will also use these terms for the corresponding modes. The spin diagonal Hamiltonian density is $\mathcal{H}_\mathrm{LL}=\sum_l \left( \mathcal{H}_\textrm{LL}^{g,l} + \mathcal{H}_\textrm{LL}^{e,l} + \mathcal{H}_\textrm{LL}^{ge,l} \right) $, where $l$ denotes the new quantum number in the spin space. The intraorbital part acting on the $\alpha=g$, and $e$ orbital state is:
\begin{equation}
\label{eq:ham-LL-intrao}
\mathcal{H}_\textrm{LL}^{\alpha,l}  (x) =\frac{\hbar}{\pi^2} u_{\alpha l}  \Big[  \frac{1}{K_{\alpha l}} \big(\partial_x \phi_{\alpha l}\big)^2  + K_{\alpha l} \big(\partial_x \theta_{\alpha l}\big)^2  \Big],
\end{equation}
and the interorbital part has the form:
\begin{equation}
\label{eq:ham-LL-intero}
\mathcal{H}_\textrm{LL}^{ge,l}  (x) =\frac{\hbar}{\pi^2} g_{l}^{ge}  \Big[ \partial_x \phi_{gl}\partial_x \phi_{el} -  \partial_x \theta_{gl}\partial_x \theta_{el}  \Big].
\end{equation}
Due to the SU(N) symmetry in the spin space the Luttinger parameters $K_{\alpha l}$, the velocities $u_{\alpha l}$ and the new couplings $g_l^{ge}$  differ only for $l=0$ and $l\neq 0$. Accordingly, the Luttinger parameters are $K_{\alpha 0}= \sqrt{\frac{2\pi\hbar v_\alpha - (N-1) g_\alpha}{2\pi\hbar v_\alpha + (N-1) g_\alpha}}$, and $K_{\alpha l}= \sqrt{\frac{2\pi\hbar v_\alpha + g_\alpha}{2\pi\hbar v_\alpha - g_\alpha}}$ for $l\neq 0$, the velocities are $u_{\alpha 0}= \sqrt{(2\pi\hbar v_\alpha)^2 - (N-1)^2 g_\alpha^2}$, and $u_{\alpha l}= \sqrt{(2\pi\hbar v_\alpha)^2 - g_\alpha^2}$ for $l\neq 0$, and finally the couplings read as $2 g_0^{ge} =  (N-1)g_{ge}^+ + (N+1) g_{ge}^- $,  and $2 g_l^{ge}= -(g_{ge}^+ - g_{ge}^-)$ for $l\neq 0$.  The interorbital part in Eq.~\eqref{eq:ham-LL-intero} mixes the two orbital states, therefore, in order to diagonalize the Hamiltonian one needs to introduce new fields as the linear combinations of the pure orbital states: 
\begin{subequations}
\label{eq:orbital-mixing}
\begin{flalign}
\Phi_{\pm ,l}= & \, \, \frac{1}{\sqrt{u_{gl}+u_{el}}}(\tilde{\phi}_{gl}\pm\tilde{\phi}_{el}), \\
\Theta_{\pm ,l}= & \, \, \frac{1}{\sqrt{u_{gl}+u_{el}}}(\tilde{\theta}_{gl}\pm\tilde{\theta}_{el}),
\end{flalign}
\end{subequations}
where we use the scaled fields $\tilde{\phi}_{\alpha l}=\sqrt{u_{\alpha l}/K_{\alpha l}}\phi_{\alpha l}$ and $ \tilde{\theta}_{\alpha l}=\sqrt{u_{ \alpha l} K_{\alpha l}} \theta_{\alpha l}$.  With these fields in Eq.~\eqref{eq:ham-LL-intero} the following scaled couplings appear: $g_{l(\tilde{\phi})}^{ge} = g_l^{ge} \sqrt{K_{gl}K_{el}/u_{gl}u_{el}}$, and $g_{l(\tilde{\theta})}^{ge} = g_l^{ge}/ \sqrt{K_{gl}K_{el}u_{gl}u_{el}}$. Now, the completely diagonal form of the Luttinger liquid part of the Hamiltonian density is
\begin{equation}
\mathcal{H}_\textrm{LL} (x) =\frac{\hbar}{\pi^2} \sum_{l,p=\pm} u_{p,l}  \Big[  \frac{1}{K_{p,l}} \big(\partial_x \Phi_{p,l}\big)^2  + K_{p,l} \big(\partial_x \Theta_{p,l}\big)^2  \Big] 
\end{equation}
with the Luttinger parameters $K_{\pm,l} = \sqrt{\frac{1\mp g_{l(\tilde{\theta})}^{ge}}{1 \pm g_{l(\tilde{\phi})}^{ge}}}$, and the velocities $u_{\pm,l} = (u_{gl} + u_{el}) \sqrt{(1\pm g_{l(\tilde{\phi})}^{ge}) (1\mp g_{l(\tilde{\theta})}^{ge})}$. In this high symmetric multicomponent Luttinger liquid state 4 velocities characterize the system. Due to the SU(N) symmetry in the spin space all the N$-1$ spin modes are degenerated, therefore, a complete mode separation can not be observed. Instead, a spin-charge separation emerges with two distinguished charge velocities corresponding to the symmetric spin combinations of the weighted mixed orbital states, and two distinguished spin-like velocities corresponding to spin combinations that are orthogonal to the previous two.

The Gaussian fixed point of the Luttinger liquid state has an extended attractive region. Nevertheless, we premise here, that with the numerical analysis of the RG equations for $^{87}$Sr isotopes, we found that the trajectories always avoid this fully gapless fixed point. Therefore, this multicomponent Luttinger liquid phase can not be realized with the two-orbital $^{87}$Sr atoms. 

\subsection{Gapped states}

The relevance of any of the processes that does not preserve the spin and charge at each branch of the spectra separately, opens one or more gaps in the excitation spectrum. The dominant fluctuations in the gapped system can be studied starting from the bosonized form of the non-Gaussian part of the Hamiltonian density. The intraorbital part for the orbit $\alpha$ is:
\begin{equation}
\label{eq:ham-int-m-bos}
\frac{1}{4\pi^2}  g_{1\perp}^\alpha  \sum_{\sigma\neq \sigma'}  \mathrm{cos} [2(\phi_{\alpha\sigma} - \phi_{\alpha\sigma'}) ],
\end{equation}
where $g_{1\perp}^\alpha = g_\alpha$. This term with relevant $g_{1\perp}^{\alpha}$ coupling pins the fields $\phi_{\alpha\sigma} - \phi_{\alpha\sigma'}$ for all unequal  $(\sigma,\sigma')$ pairs, therefore  all the spin-like $\phi_\alpha$ modes become gapped.  The interorbital density-density interaction has similar form:
\begin{equation}
\label{eq:ham-int-ge-bos}
\frac{1}{4\pi^2} \sum_{\sigma , \sigma'} \left[ g_{1\parallel}^{ge} \delta_{\sigma,\sigma'} + g_{1\perp}^{ge} (1-\delta_{\sigma,\sigma'})\right]  \mathrm{cos} [ 2(\phi_{g\sigma} - \phi_{e\sigma'})],
\end{equation}
where $g_{1\parallel}^{ge} = g_{ge}$, and $g_{1\perp}^{ge} = g_{ge}$. The two terms can be relevant or irrelevant  independently of each other. The $g_{1\parallel}^{ge}$ term opens a gap in the charge as well as all the spin modes of the $\phi_{g}-\phi_{e}$ fields (i.e. their antisymmetric combination in the orbital states), while with relevant $g_{1\perp}^{ge}$ term only the spin sector of the $\phi_g-\phi_e$ becomes gapped, and the charge mode remains free. Note, that in principle, if $g_{1\parallel}^{ge} + g_{1\perp}^{ge}$ scales to zero, the spin sector remains gapless, but in case of $^{87}$Sr we did not find such a fixed point, either. Finally the interorbital exchange is:
\begin{multline}
\label{eq:ham-int-ge-ex-bos}
\frac{1}{4\pi^2}   \sum_{\sigma , \sigma'} \Big\{ g_{1\parallel}^{ge} \delta_{\sigma,\sigma'}   \mathrm{cos} [2(\phi_{g\sigma} - \phi_{e\sigma})]  \\ 
+ \tilde{g}_{1\perp}^{ge} (1-\delta_{\sigma,\sigma'}) \Big[ (2 \mathrm{cos}^2\phi_1  -1 )(2 \mathrm{cos}^2\theta_1 -1) - \mathrm{sin}\phi_1  \mathrm{sin}\theta_1  \Big] \\
+ \tilde{g}_{2\perp}^{ge} (1-\delta_{\sigma,\sigma'}) \Big[ (2 \mathrm{cos}^2\phi_2  -1 )(2 \mathrm{cos}^2\theta_2 -1) - \mathrm{sin}\phi_2  \mathrm{sin}\theta_2  \Big] \Big\}
\end{multline}
where $g_{1\parallel}^{ge} = -g_{ge}^\mathrm{ex}$, while  $\tilde{g}_{\perp}^{ge}=g_{ge}^\mathrm{ex}$ and $\tilde{g}_{2\perp}^{ge} = g_{ge}^\mathrm{ex}$, and the short hand notations have been introduced: $\phi_1=\phi_{g\sigma}-\phi_{g\sigma'} +\phi_{e\sigma}-\phi_{e\sigma'}$, and $\phi_2 = \phi_{g\sigma}+\phi_{g\sigma'} - \phi_{e\sigma}-\phi_{e\sigma'}$, respectively, and the identical combinations of the dual fields. Again, the relevant $g_{1\parallel}^{ge}$ term pins the $\phi_{g}-\phi_e$ fields in the whole spin space and makes the corresponding charge and spin modes gapped. The other two terms do not affect on the $\phi$ fields only, but their dual fields $\theta$, too. The effect of the $\tilde{g}_{2\perp}^{ge}$ term on the $\phi$ fields is the same as that is of the $g_{1\parallel}^{ge}$ term. Contrary, the
$\tilde{g}_{1\perp}^{ge}$ term pins the symmetric combination in the orbital state, and antisymmetric in the spin state, therefore the spin sector of the $\phi_g + \phi_e$ fields becomes fully gapped, while the corresponding charge mode can fluctuate freely. On the $\theta$ fields the $\tilde{g}_{1\perp}^{ge}$ and $\tilde{g}_{2\perp}^{ge}$ terms take the same effect as they do on the $\phi$ phase fields. $\tilde{g}_{1\perp}^{ge}$ pins the antisymmetric combinations in both the orbital and the spin states, i.e. the  spin sector of the $\theta_g + \theta_e$ fields becomes fully gapped. With relevant $\tilde{g}_{2\perp}^{ge}$ coupling all the orbital-antisymmetric combination of the dual fields are pinned, therefore the charge and spin sector of the $\theta_g - \theta_e$ are gapped.

\section{Possible phases of $^{87}\mathrm{Sr}$ atoms}
\label{sec:Sr}

With the analysis of the RG equations the relevant interactions can be determined and taking into account their effect on the various modes, the possible phases of the system can be studied. 
However, a complete analysis of the four dimensional parameter space ($g_g$, $g_e$, $g_{ge}^+$, $g_{ge}^-$) would be quite difficult. 
For $^{87}$Sr the ground state scattering length is known $a_g = 96.2 a_0$ (where $a_0\approx 0.053$ nm is the Bohr radius) \cite{Escobar2008} and there exists an estimation  for the value of the scattering  length $a_{ge}^- \approx -300a_0$ \cite{Campbell2009}. Therefore, we can fix the corresponding two coupling constants $g_g$ and $g_{ge}^-$,  and only the two-dimensional parameter space of the couplings $g_e$ and $g_{ge}^+$ remains to investigate. We hope that soon there will be available the various scattering length for further atoms/isotopes, too. From now we focus on the possible phases of the $^{87}$Sr isotope. The total electron angular momentum of the Strontium-87 is 0 and its nuclear spin is 9/2. Accordingly, the model defined by Eqs.~\eqref{eq:ham-kin}-\eqref{eq:ham-int-ge} has an SU(10) symmetry.

\begin{figure}
\includegraphics[scale=0.38]{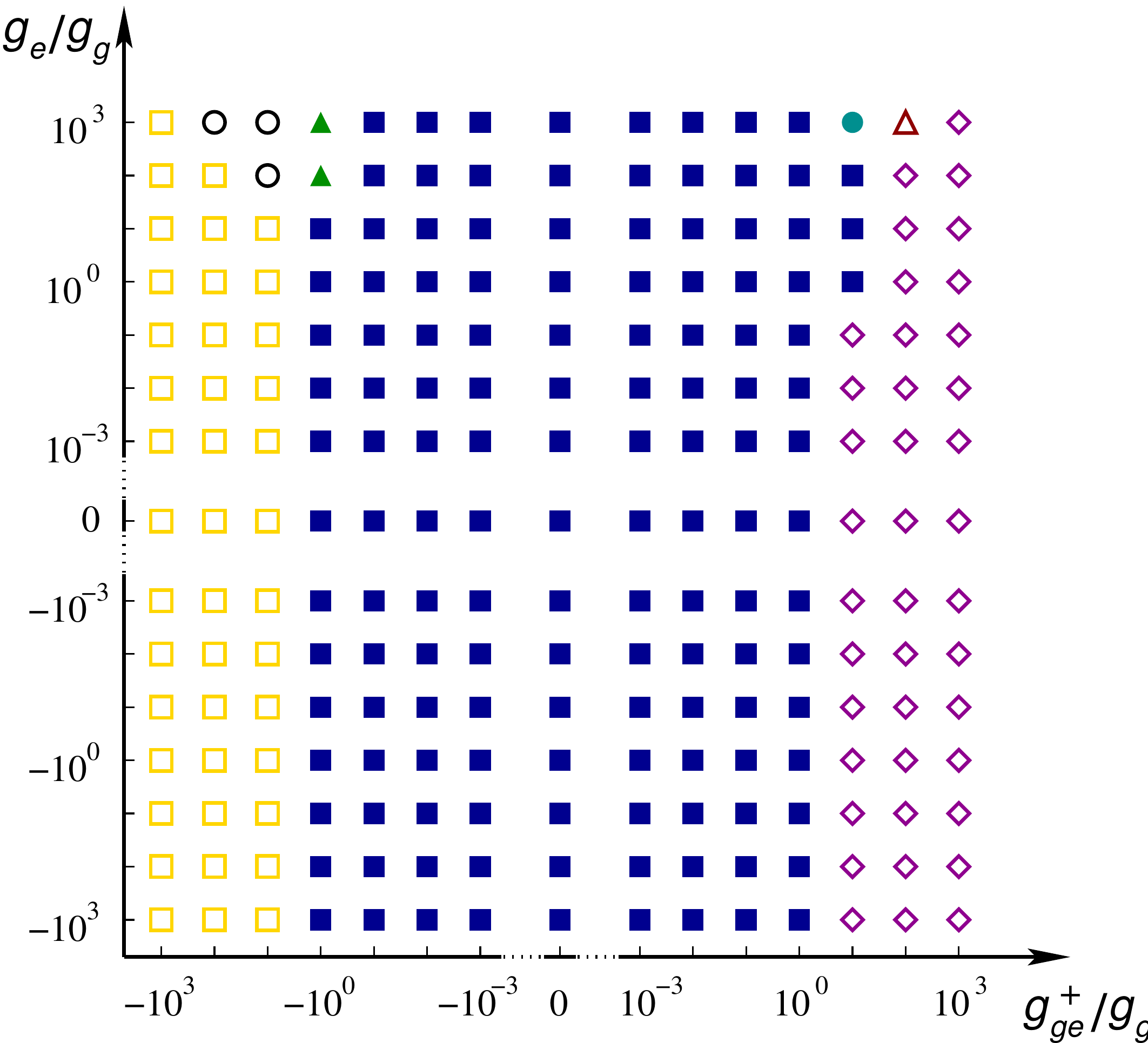}
\caption{(Color online) The fixed point structure of the $^{87}$Sr isotope on the plane $(g_{ge}^+,g_e)$ settled by the value $g_{ge}^-/g_g=-3$. The different symbols related to different fixed point of the RG trajectories, the definition of the various symbols can be found in Table \ref{tab:fixed-points}. Note that on both the horizontal and vertical axes we used logarithmic scale. }
\label{fig:phase-diagram}
\end{figure}

\begin{table}[b!]
\begin{tabular}{| c | c | c | c | c | c | c | c | c | c |  }
\hline
~ & $\square$ & $\blacksquare$  & $\Diamond$ & $\vartriangle$ & $\blacktriangle$ & {\Large $\circ$} & {\Large $\bullet$}  \\ 
\hline
$g_{1\perp}^g$ & $-\infty$ &                  $-\infty$   & $-\infty$ & $-\infty$ & $+\infty$ & $+\infty$ & $+\infty$ \\ 
$g_{1\perp}^e$ & $-\infty$ &                  $-\infty$   & $-\infty$ & $-\infty$ & $+\infty$ & $+\infty$ &  $+\infty$ \\
$g_{1\parallel}^{ge}$ & $-\infty$ &        $-\infty$  & $+\infty$ & $-\infty$ & $+\infty$ & $+\infty$ & $-\infty$ \\ 
$g_{1\perp}^{ge}$ & $-\infty$ &            $-\infty$  & $+\infty$ & $-*$ & $-*$ & $-*$ & $+*$ \\
$\tilde{g}_{1\perp}^{ge}$ & $-*$ &  $+*$ & $+*$ & $0$ & $0$ & $0$ & $0$ \\ 
$\tilde{g}_{2\perp}^{ge}$ & $-*$ & $+*$ & $+*$ & $+*$ & $+*$ & $-*$ & $+*$\\
\hline
\end{tabular}
\caption{The definition of the fixed points of the RG trajectories (see also Fig. \ref{fig:phase-diagram}). $\pm*$ denote various finite or even infinite positive or negative fixed point values, their absolute values depend on the initial values of the couplings. }
\label{tab:fixed-points}
\end{table}

We have analyzed numerically the RG equations \eqref{eq:betafunctions} with the initial values \eqref{eq:RG-initial_values}. As initial values we took $g_g$ as unit, and $g_{ge}^- /g_g=-3$ that is reliable in the precision of the estimation. Since the scattering length can take any values in a wide range \cite{Escobar2008}, and even their sign can differ, we carried out the analysis in a range where the remaining two couplings $g_e$ and $g_{ge}^+$ can be smaller or larger with 3 order of magnitude than $g_g$. The basis of the phase diagram provided by the fixed point structure is given in Fig.~\ref{fig:phase-diagram}. For the better visibility we have used logarithmic scale on the axes, and the meanings of the symbols are listed in Table~\ref{tab:fixed-points}. As we have seen the interaction terms that scale to the strong coupling regime pin various bosonic fields and the remaining free fields determine the dominant fluctuations in the system. From this point of view the scattering processes scaling to the infinity or to a large finite value affect similar way, therefore we do not distinguish them.

According to the above analysis with respect to the effect of the various interaction terms on the fields $\phi$ and $\theta$, one can recognize that the spin sector of the $\phi$ fields is fully gapped in the whole $(g_{ge}^+,g_e^{\phantom{\dagger}})$ plane because of the always relevant $g_{1\perp}^g$, $g_{1\perp}^g$, and $g_{1\perp}^{ge}$ terms. And similarly, $g_{1\parallel}^{ge}$ always scales to the strong coupling, therefore the charge mode of the anti-bonding (orbital antisymmetric) $\phi_g-\phi_e$ field is also gapped and only the charge mode of the bonding (symmetric combination in the orbital states) $\phi_g + \phi_e$ field remains free. The dynamics of the $\theta$ fields is determined by the orbital exchange terms $\tilde{g}_{1\perp}^{ge}$ and $\tilde{g}_{2\perp}^{ge}$. $\tilde{g}_{2\perp}^{ge}$ is always relevant, therefore the charge and the spin-like combinations of the anti-bonding $\theta_g-\theta_e$ dual fields are pinned leaving to fluctuate freely only the bonding dual field combinations. Additionally, the $\tilde{g}_{1\perp}^{ge}$ coupling also relevant in the largest part of the phase diagram, that pins the orbital-symmetric combinations in the whole spin sector, and only the charge mode of the dual fields $\theta_{g0} + \theta_{e0}$ remains free. Accordingly, in these phases the spin degrees of freedom are frozen out and the low energy physics of the system is equivalent with the one of a two-orbital spinless fermion system that is not affected by the underlying (gapped) spin order.

\subsection{Incommensurate fillings}

Let us first consider the case when there is no relevant umklapp processes. On the largest part of the phase diagram in Fig. \ref{fig:phase-diagram} the $\tilde{g}_{1\perp}^{ge}$ coupling is relevant, therefore the dominant fluctuations are determined by only the charge combinations $\phi_{g0} + \phi_{e0}$ and $\theta_{g0} + \theta_{e0}$. The $2k_\textrm{F}$ density-waves fluctuate with $\mathcal{O}_{2k_\textrm{F}\textrm{-DW}} \sim \textrm{e}^{\textrm{i}(\phi_{g0} + \phi_{e0})}$, and applying the transformation Eqs.~\eqref{eq:orbital-mixing} one finds that its correlation function decays with the distance $r$ as 
$|r|^{-\Delta_{\Phi_{+0}}-\Delta_{\Phi_{-0}}}$, with exponent
\begin{equation}
\label{eq:d_phi_l}
\Delta_{\Phi_{\pm l}} = \frac{1}{4\pi} \left[ \frac12 \frac{  \sqrt{ \frac{ K_{gl} }{ u_{gl} } } \pm \sqrt{ \frac{ K_{el} }{ u_{el} } } }{ \sqrt{ 1 \pm g_l^{ge} \sqrt{ \frac{ K_{gl}K_{el} }{ u_{gl}u_{el} } } } } \right]^2,
\end{equation}
and with $l=0$. Note, that the $4k_\textrm{F}$ density-waves fluctuate with $\mathcal{O}_{4k_\textrm{F}\textrm{-DW}} \sim \textrm{e}^{\textrm{i}2(\phi_{g0} + \phi_{e0})}$, therefore, they are always suppressed by the $2k_\textrm{F}$ quasi-long-range density oscillations. 

\begin{figure}[t]
\includegraphics[scale=0.25]{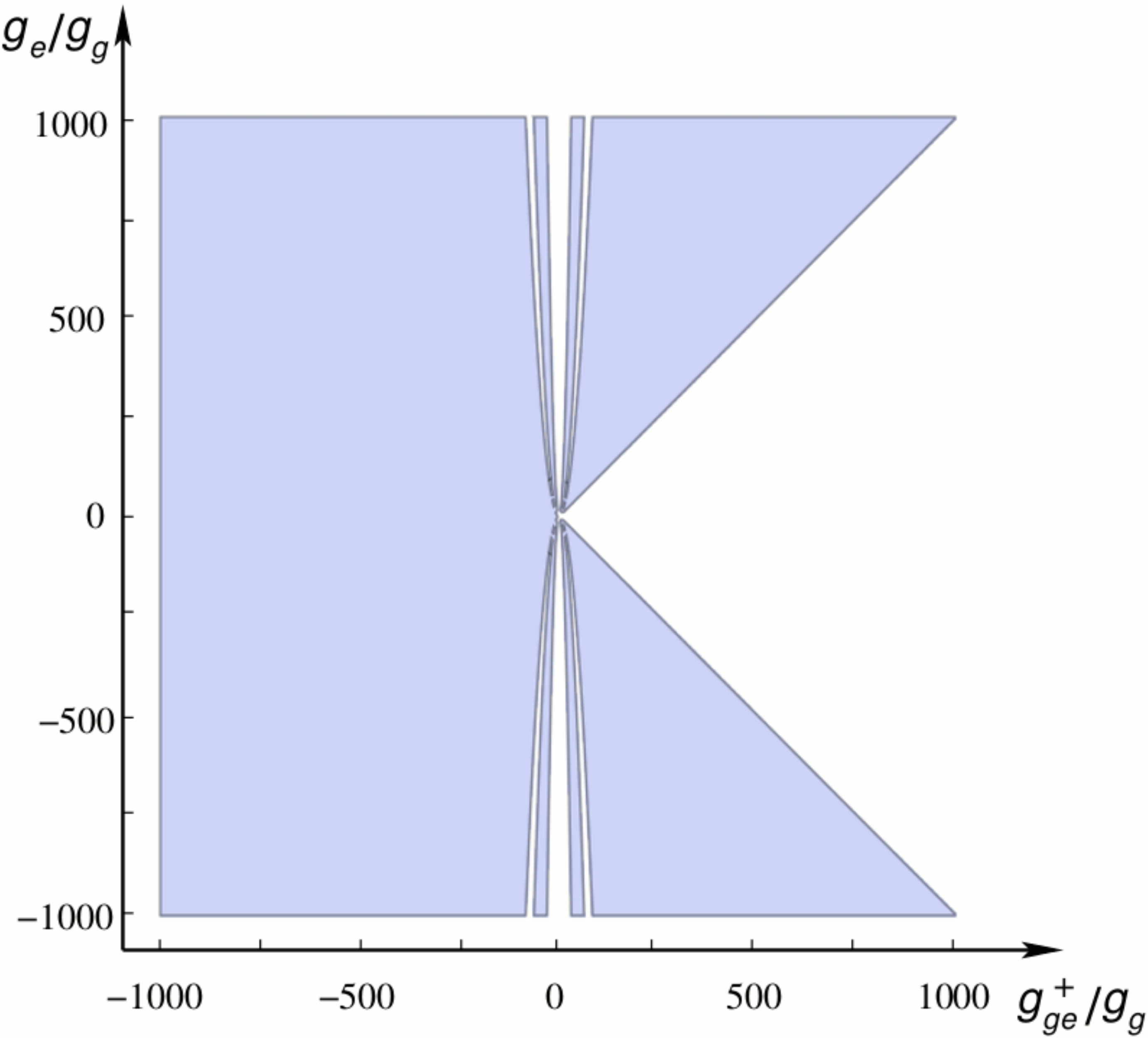}
\caption{(Color online) The phase diagram of the $^{87}$Sr isotope on the plane $g_{ge}^-/g_g =-3$. The dark (blue) region shows the parameter regime where the density fluctuations dominates, while in the white regions the superconducting instabilities show slowest decay. 
}
\label{fig:phase-diagram-1}
\end{figure}

\begin{figure}[b]
\includegraphics[scale=0.25]{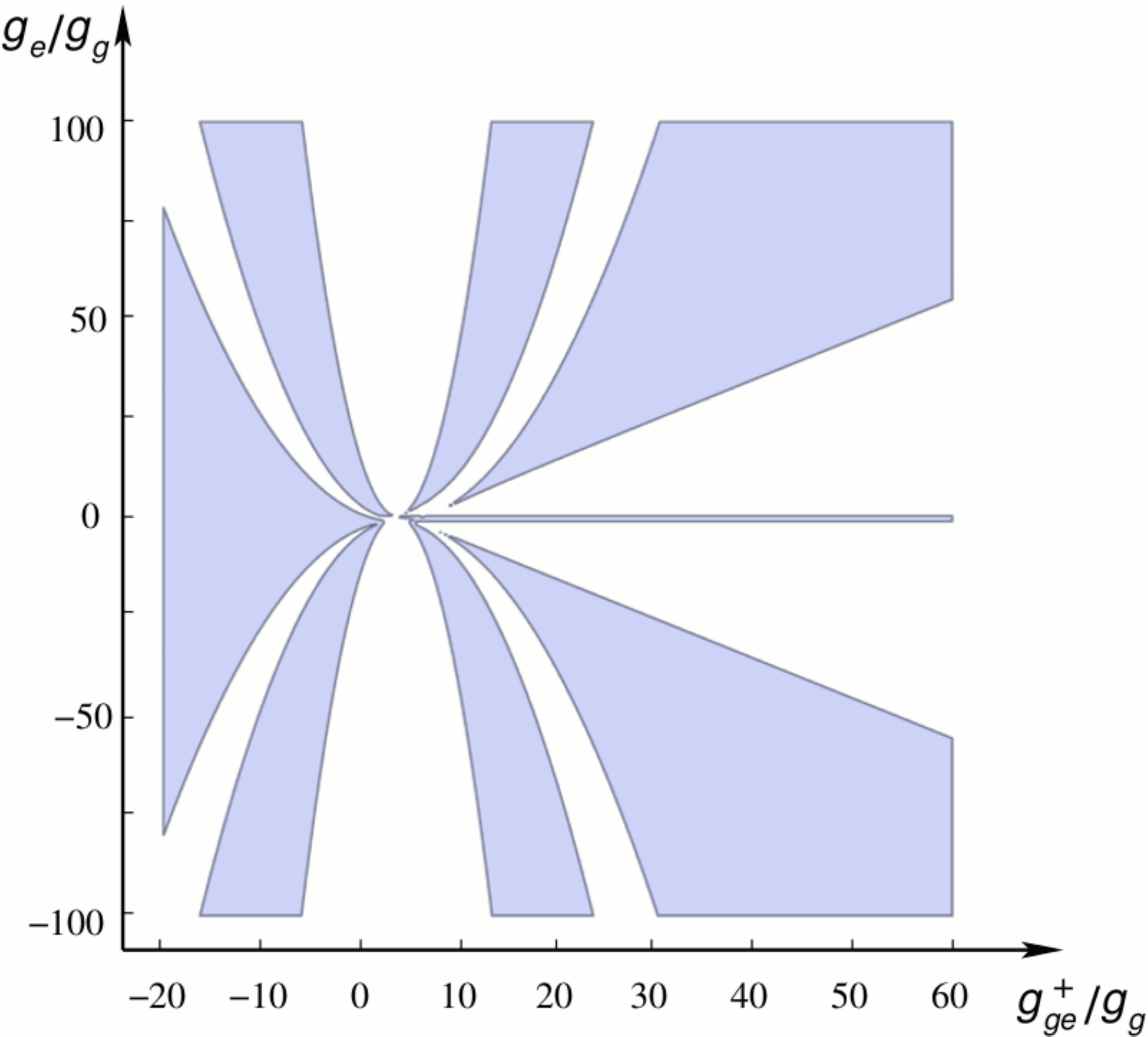}
\caption{(Color online) A zoom of the phase diagram in  Fig.~\ref{fig:phase-diagram-1} to the moderated values of the interactions is presented here in order to get better visibility of the structure of the phase boundaries. }
\label{fig:phase-diagram-2}
\end{figure}

Nevertheless, the Cooper pair instabilities are characterized by $\mathcal{O}_\textrm{SC} \sim \textrm{e}^{\textrm{i}(\theta_{g0} + \theta_{e0})}$, that can win over the  $2k_\textrm{F}$ density-fluctuations. As it was concluded above, the spin degree of freedom is frozen out in this part of the phase diagram (namely, where $\tilde{g}_{1\perp}^{ge}$ is relevant). Therefore, only two different Cooper pairs can be distinguished: the orbital singlet, and the orbital triplet pairs. The decay of both Cooper pair correlation functions are found to be determined by the same phase field oscillation due to the numerous pinned fields. The correlation function of the Cooper pairs decays as $|r|^{-\Delta_{\Theta_{+0}}-\Delta_{\Theta_{-0}}}$,  where
\begin{equation}
\label{eq:d_theta_l}
\Delta_{\Theta_{\pm l}} = \frac{1}{4\pi} \left[ \frac12  \frac{  \frac{1}{ \sqrt{ K_{gl} u_{gl} } } \pm  \frac{1}{ \sqrt{ K_{el} u_{el} } } }{ \sqrt{ 1 \pm g_l^{ge} \frac{1}{ \sqrt{ K_{gl}K_{el} u_{gl}u_{el} } } }} \right]^2 ,
\end{equation}
and $l=0$. Therefore, if $\Delta_{\Theta_{+0}}+\Delta_{\Theta_{-0}} < \Delta_{\Phi_{+0}} + \Delta_{\Phi_{-0}}$, the superconducting instability dominates. 
In Fig.~\ref{fig:phase-diagram-1} we plotted the sign of the quantity $ \Delta_{\Phi_{+0}} + \Delta_{\Phi_{-0}} - \Delta_{\Theta_{+0}} - \Delta_{\Theta_{-0}}$. Where it is positive, the Cooper pair correlations show slower decay, the dominant instability is the pair fluctuations (white region). Otherwise, a density waves like quasi-long-range order characterizes the system with $2k_\mathrm{F}$ periodicity (blue region). Note, that in this case we used linear scale instead of the logarithmic scale used in case of Fig. \ref{fig:phase-diagram}, in order to emphasize the nontrivial structure of the phase diagram: for intermediate values of the coupling $g_{ge}^+$ the phase boundary between the superconducting and density wave state has a rather complicate structure that is shown in Fig.~\ref{fig:phase-diagram-2}.  The shape of the phase boundaries is not sensitive qualitatively to the value of $g_{ge}^-/g_g$ at least as long as it is in the order of 10. 
The Luttinger parameters and accordingly the scaling dimensions in Eqs.~\eqref{eq:d_phi_l}, and \eqref{eq:d_theta_l} depend on the couplings of the various interactions via a complicated square root function, that leads to a complex structure of the phase diagram.

In order to understand more deeply the emerging phase boundaries let us have a look at a segment of the phase diagram setting by a fixed value of $g_e/g_g$. A few alternation of the density wave and the superconducting state can be observed when $g_{ge}^+$ small compare to the coupling $g_e$, and for dominant orbital-symmetric couplings Cooper pairing gains again over the density fluctuations. At this point is it important to emphasize that in the whole phase diagram both instabilities show algebraic decay and the only difference between the phases is the dominant fluctuation that is chosen to characterize the phase. Considering the scaling dimensions in Eqs.~\eqref{eq:d_phi_l}, and \eqref{eq:d_theta_l} it is easy to see that the two competing fluctuations are mostly driven by the weighted coupling $g_0^{ge}=[(N-1)g_{ge}^+ + (N+1) g_{ge}^-]/2 = [N g_{ge} - g_{ge}^\mathrm{ex}]/2$, i.e. the weighted density-density interaction and orbital exchange interaction. The weight of the density fluctuations relates to the bare compressibility $\kappa_\alpha$ of the particles in the ground state and in the excited state as $\kappa_{\alpha} \sim K_{\alpha 0}/u_{\alpha 0}$, with $\alpha=g$, or $e$. Similarly, the weight in case of the Cooper pair fluctuations relates to the bare conductivity $\sigma_\alpha$ in the two different orbital states as $\sigma_{\alpha} \sim K_{\alpha 0}u_{\alpha 0}$ (see e.g. in Ref.~\cite{schulz95a}). Note, that the bare orbital compressibility and conductivity are renormalized by the interorbital scatterings described by Eq.~\eqref{eq:ham-LL-intero}, and the compressibility and conductivity of the interacting system relate to the Luttinger parameters $K_{\pm,0}$, and the velocities $u_{\pm,0}$. Nevertheless, the bare orbital parameters can also be measured within an independent experiment (see Sec.~\ref{sec:experiment}). The competition of the four dressed interactions $\sqrt{\kappa_g\kappa_e} g_{ge}$, $\sqrt{\kappa_g\kappa_e} g_{ge}^\mathrm{ex}$, $g_{ge}/\sqrt{\sigma_g \sigma_e}$, and $g_{ge}^\mathrm{ex}/\sqrt{\sigma_g \sigma_e}$ produces the alternation of the phases as their relative values are changing. The competition of the dressed interactions takes place in the region where the weighted couplings are comparable, i.e. around $g_{ge}^+ \sim g_{ge}^-$, and around $g_{ge}^+ \sim g_e$.  Due to the fixed value of $g_{ge}^-/g_g=-3$, the first region is restricted to a relatively narrow interval of $g_{ge}^+$.

In certain regions of the phase diagram the $\tilde{g}_{1\perp}^{ge}$ coupling scales to zero (see Table \ref{tab:fixed-points}.), and due to its irrelevance all the spin-antisymmetric, orbital symmetric combinations of the dual fields $\theta$ can fluctuate freely. In this case the $2k_\textrm{F}$ density wave or the Cooper pairs can compete with or even be suppressed by $2k_\textrm{F}$ spin-carrier density wave --- similar to spin-density wave in the two-component case. The $2k_\textrm{F}$ spin-carrier density wave fluctuates with $\mathcal{O}_{2k_\textrm{F}\textrm{-SDW}}^{(l)} \sim \textrm{e}^{\textrm{i}(\phi_{g0} + \phi_{e0})} \textrm{e}^{\textrm{i}(\theta_{gl} + \theta_{el})/2}$, where $l=1,\ldots , 9$. Due to the SU(N) symmetry in the spin space, the scaling dimension of $\mathcal{O}_{2k_\textrm{F}\textrm{-SDW}}^{(l)}$ does not depend on $l$, the corresponding correlation functions decay as $|r|^{-\Delta_{\Phi_{+0}}-\Delta_{\Phi_{-0}} -(\Delta_{\Theta_{+l}}+\Delta_{\Theta_{-l}})/2 }$. The exponents are given by Eqs.~\eqref{eq:d_phi_l} and \eqref{eq:d_theta_l} with $l\neq 0$. Nevertheless, we found that in the parameter region where the dual field combination $\theta_{gl}+\theta_{el}$ can fluctuate freely, the spin fluctuation can not dominate over the density wave or the Cooper pair instabilities.

\subsection{Commensurate fillings}

In case of a finite lattice, in principle, incommensurate filling is not possible, since there always exist integer (and relative prime) $p$, and $q$ for which  $2k_\textrm{F}p/q = 2\pi/a$. In these cases the leading order umklapp processes describing scatterings with momentum transfer $4k_\mathrm{F}$, $6k_\mathrm{F}$, $8k_\mathrm{F}$ etc. can be relevant. These higher order umklapp processes relate to multifermion scatterings: at $p/q$ filling the leading order umklapp processes can be described by $q$-particle scatterings. However, within the applied RG procedure such multiparticle umklapp processes are never generated, at the corresponding filling they can be relevant. The bosonized form of the umklapp term consists cosines of the summation over the $q$ phase fields $\phi$ in all possible combinations (see Eq.~\eqref{eq:umklapps}).  However, the umklapp processes couples only to the symmetric combination of the $q$ fields, in general they mix all the charge and spin modes, and also the orbital-symmetric and orbital-antisymmetric modes. We have seen above that in case of $^{87}$Sr atoms at incommensurate fillings the dominant fluctuations are always determined by the symmetric combinations in the orbital degree of freedom. Therefore, the umklapps can open gap only in the spectrum of the orbital-symmetric modes, so it is reasonable to consider only them.

Therefore, we restrict our analysis to the effect on the $\phi_{g0} + \phi_{e0}$, and $\phi_{gl} + \phi_{el}$ field combinations. These terms pin the corresponding modes, and suppress the site centered $2k_\mathrm{F}$-CDW state. Instead, for positive values of the umklapp processes spin-Peierls-like bond order of the orbital-symmetric fields occur with periodicity determined by simply the relation of the filling factor and N (see Appendix~\ref{sec:umklapps}).  Accordingly, at half filling the emergence of a dimer order is expected, at third filling a similar bond order with periodicity $3a$, a so called trimerized state, and so on, as long as the filling is $p/q$ and $q<\,$N. At 1/N-filling, the umklapps couple only to the charge modes of the orbital-symmetric combination of the phase fields, in general the spin modes would remain gapless, and a homogeneous ground state would be expected. However, for Strontium-87, due to the relevant backward scatterings the spin modes are gapped anyway. Therefore, at 1/10-filling, too, spin-Peierls-like bond order of the orbital-symmetric fields emerges with periodicity $10a$.

\section{Experimental aspects}
\label{sec:experiment}

Finally, we discuss some perspectives and challenges of the experimental probing of the various phases. There are two relevant questions: how to study the low-energy excitations, and how to probe the emerging (quasi-long range) pairing and density wave states. There are several methods to support the occurrence of the superfluid or the density wave orders, usually based on the measurement of the one-particle excitation gap, or the momentum distribution \cite{bloch08a,ketterle08a}. In principle, the latter method would be especially effective in case of Cooper-like pairs consisting one particle in the electronic ground state and another in the excited state. These pairs have finite momentum in the order of the difference of the two Fermi momentums: $\pm(k_\mathrm{F}^g-k_\mathrm{F}^e)$. Accordingly, the emerging superfluid state is analogous with the celebrated Fulde-Ferrell-Larkin-Ovchinnikov state. FFLO states have been studied extensively in various spin- and mass-imbalanced ultracold atomic systems, and their experimental realization is in progress \cite{Partridge06a,*Zwierlein06a}, even in the one-dimensional case \cite{Liao10a}. 

In our case, it is not obvious how to find the suitable excitations coupled to the complicated collective modes that characterize the low-energy behavior (see Eq.~\eqref{eq:orbital-mixing}).
The bosonic fields introduced in Eq.~\eqref{eq:boson_fields} relate to the density as $-\partial_x \phi_{\alpha,\sigma}(x) /\pi = n_{\alpha,\sigma} (x)$, so the fluctuations of the bosonic fields correspond to density oscillations. The low energy excitations of the system are characterized by these density fluctuations, and the corresponding boson fields occur only in the Gaussian part of the Hamiltonian. Accordingly, their excitation spectrum is sound-like: $\hbar \omega_{\alpha l} = u_{\alpha l}q$, where the $u_{\alpha l}$ sound velocity has been defined in Sec.~\ref{sec:LL}.  We have seen that the $\phi_{\alpha l}$ fields are the linear combinations of the $\phi_{\alpha,\sigma}$ fields, so they still relate to density fluctuations, and without interorbital interaction they are the eigenmodes of the system. Within two independent measurements, one performed with an atomic cloud that contains atoms in the electronic ground state, and another one with atoms in the electronic excited state, the sound velocities $u_{gl}$, and $u_{el}$ can be determined by an external perturbation (excitation) coupled to the corresponding mode. For instance, Bragg spectroscopy provides an effective tool to study the low-energy density  excitations, and determine the sound velocity. Note, that in general a spin- and orbital-selective method is desired to determine all the $u_{\alpha l}$ velocities. When the atomic cloud consists atoms in the electronic ground state as well as in the excited state, the interorbital couplings become relevant. The interorbital coupling in Eq.~\eqref{eq:ham-LL-intero} has two effects. On one hand it changes the eigenmodes to $\Phi_{\pm,l}$ that are difficult to probe since it is not obvious how to excite them directly. On the other hand the interorbital coupling renormalizes the $u_{\alpha l}$ sound velocities. The renormalized velocities can be measured in this case too, by exciting the corresponding density modes $n_{\alpha,l}$. We have seen that in case of the Sr-87 isotope only the $\phi_{g0} + \phi_{e0}$ phase field combination can fluctuate freely. Fortunately, as a symmetric combination in both the spin and orbital degrees of freedom, it describes an orbital bounding charge mode, i.e. relates to the total density of the system that can be probed by Bragg spectroscopy.

\section{Conclusions}

In this work we considered a high spin SU(N) symmetric fermionic system confined in a one-dimensional chain, and analyzed the possible consequences of the relevance of an additional degree of freedom with two possible internal states. Such an additional two-state degree of freedom can be realized as the ground state and the first excited electronic state of the atoms. The corresponding Hamiltonian and the Hilbert space are analogous to a two-orbital system providing a prefect candidate to mimic the physics of two-orbital systems \cite{gorshkov10a}. 

The $\beta$-functions of the renormalization group transformation have been determined up to one-loop order in the most general case, i.e. general spin dependence was assumed for the scattering processes. The equations contain the SU(N) symmetric case as a special case. With the help of Eqs~\eqref{eq:betafunctions} the renormalization flows of two-orbital systems with arbitrary spin depending two-particle interactions were determined. We have diagonalized the quadratic part of the Hamiltonian that describes a 2N-component Luttinger liquid. The spin sector of this multicomponent Luttinger liquid state is highly degenerated due to the SU(N) symmetry in the spin space. As a consequence of this degeneracy the Luttinger liquid state is characterized by 4 velocities. 

We applied the analysis to determine the phase diagram of the $^{87}$Sr isotope that can be considered as a potential candidate to realize experimentally a two-orbital high-spin system. The $^{87}$Sr isotopes have closed electronic outer shell, and have $F=9/2$ hyperfine spin, therefore in principle an effective SU(10) symmetric system can be modeled by them. We concluded that the 20-component Luttinger liquid state is absent from its phase diagram. We found that there exist different nonquadratic, gapped fixed points related to dominant density fluctuation or superconducting instability, depending on the values of the couplings. The phase boundary between the pair and the density fluctuating states has a nontrivial shell structure for moderate values of the interactions according to the competition of various weighted interorbital interactions. The experimental probe of the above presented nontrivial phase structure would be very desired, as a new probe of the hydrodynamic treatment of one-dimensional quantum liquids. Additionally, the system provides a possible realization of an alternative FFLO state where the finite momentum of the pairs comes from the difference of the Fermi momentums in the two orbital states, instead of spin imbalance or bare mass difference between the interacting particles.

\section*{Acknowledgements}

This work was supported by the National Research Found (OTKA) No. K105149 and K100908.

\appendix

\section{Derivation of the RG equations}
\label{sec:app-0}

In this Appendix we summarize some details of the calculations of the $\beta$-functions \eqref{eq:betafunctions}. The theoretical background of the RG treatment can be found in several textbooks and reviews (see e.g. Ref.~\cite{solyom79a, gogolin04a,giamarchi04a}), here we give only the problem-specific details of the corresponding vertex corrections and the calculation of their contributions.

\begin{figure}[t]
\includegraphics[scale=0.33]{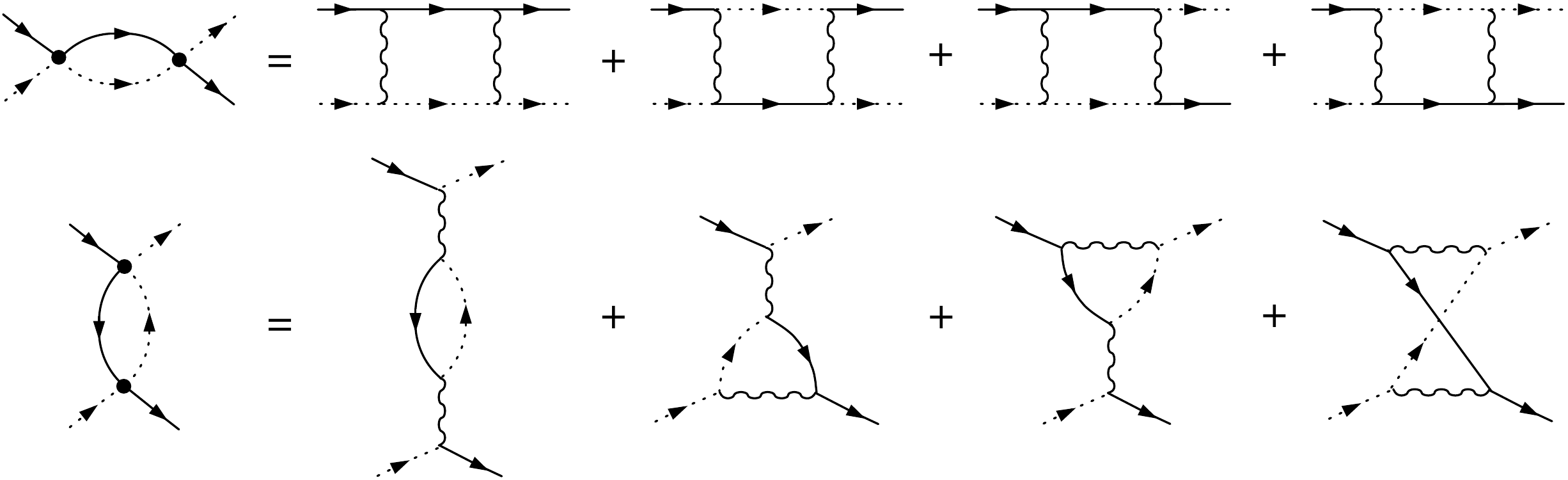}
\caption{The vertex corrections up to one-loop order. The vertices with the same structure can be classified into two different channel: the upper line shows the contributions of the Cooper channel, the lower line shows that of the zero-sound channel. The solid lines refer to a right moving particle, the dashed lines refer to a left moving one, and the wavy lines denote the interaction. Note, that in the short hand notation of the vertices the interactions are denoted by a black dot.
}
\label{fig:vertices}
\end{figure}

The applied RG transformation based on the perturbation calculation of the vertex corrections up to the leading order, that is in our case the one-loop order. Away from half-filling there is no umklapp processes, the only contributing vertices are listed in Fig.~\ref{fig:vertices}. In the first line such one-loop order corrections are collected where a particle pair is propagating in the intermediate state. This is the so called Cooper channel. In the second line in the intermediate state a particle-hole pair is propagating, this is the zero sound channel. The corresponding vertices are logarithmically divergent, and the vertices with the same structure differs only (apart from the bare couplings) in a numerical factor. This factor comes form the Feynman rules of the given problem: summation over spin and a $-1$ sign for each loop. Apart from these factors, the contribution of the Cooper channel is:
\begin{equation}
\Gamma_\mathrm{Cooper} \sim -\frac{1}{\pi(v_\alpha +v_{\alpha'})} \left( \ln \left| \frac{\omega}{E_0}\right| - \mathrm{i}\frac{\pi}{2} \right)
\end{equation}
where $\alpha$ and $\alpha'$ refer to the orbital state of the two particles in the intermediate state. Note, that for simplicity we used only one frequency parameter $\omega$, and similarly one band-width cut-off $E_0$ to determine the contribution of the vertices. In the zero sound channel, the logarithmic is very similar, differs only in its sign:
\begin{equation}
\Gamma_\mathrm{z-s} \sim \frac{1}{\pi(v_\alpha +v_{\alpha'})} \left( \ln \left| \frac{\omega}{E_0}\right| - \mathrm{i}\frac{\pi}{2} \right).
\end{equation}

\begin{figure}[t]
\centering
\includegraphics[scale=0.23]{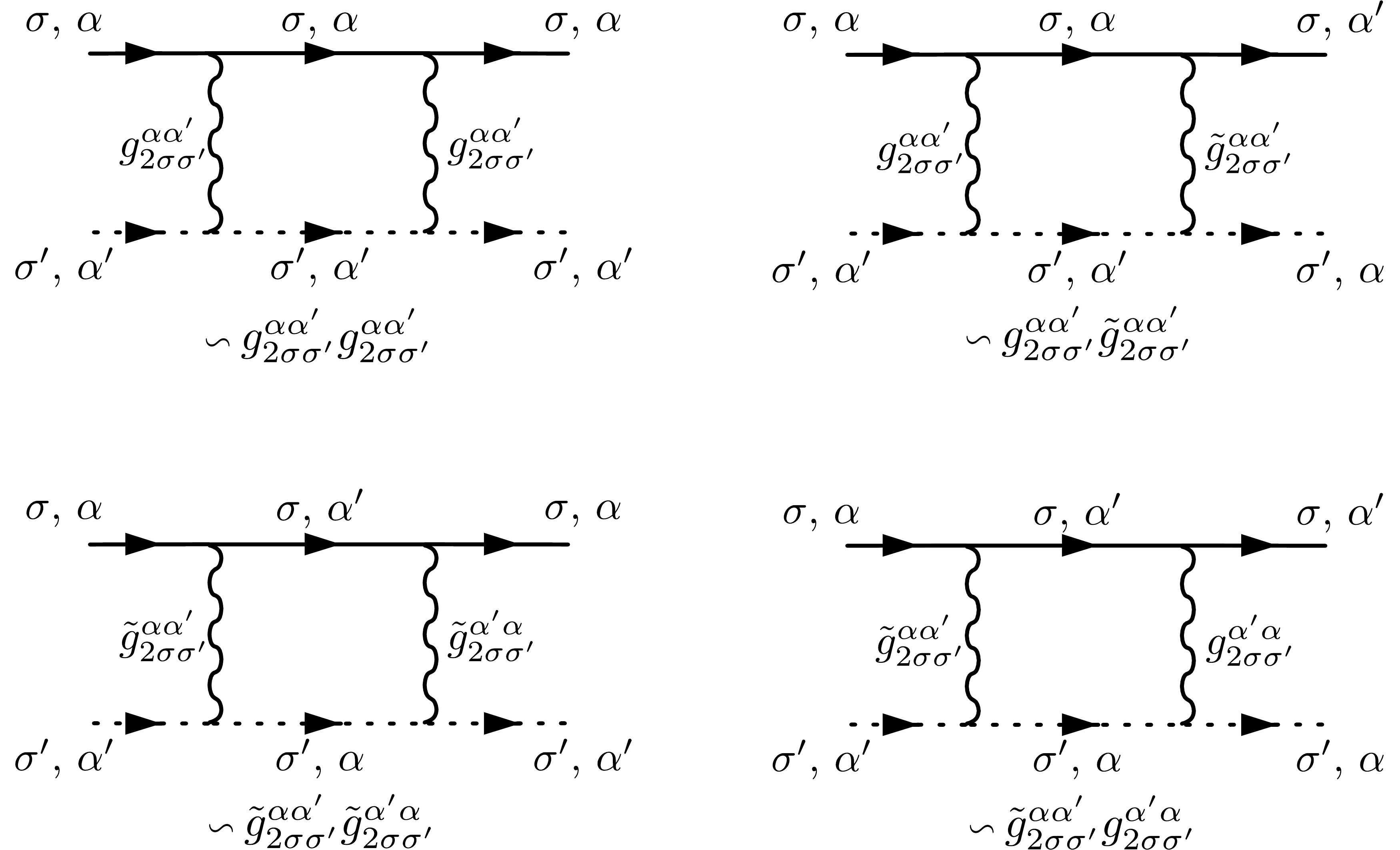}
\caption{The possible spin and orbital configurations for the first type of the Cooper channel vertices in Fig.~\ref{fig:vertices}. 
}
\label{fig:vertex-example}
\end{figure}

As an illustration let us consider the first vertex of the Cooper channel in Fig.~\ref{fig:vertices}. The possible processes are shown in Fig.~\ref{fig:vertex-example}.  This vertex gives contribution to the small momentum transfer processes with $q\approx \pm(k_\mathrm{F}^{\alpha_{ir}} - k_\mathrm{F}^{\alpha_{il}})$. Depending on the relative value of the orbital indices of the initial and final state, they can relate to the orbital-flipping process or the one that preserves the orbital state: the vertices in the column on the left hand side relate to the correction of $\Gamma_2$, and the ones in the right column relate to $\tilde{\Gamma}_2$. For the other vertices the possible spin and orbital configurations can be determined similarly. Exploiting the symmetry properties of the bare vertices in their spin and orbital indices, and collecting the contributions of every one-loop order corrections one can arrive to the following expression of the vertices:
\begin{subequations}
\label{eq:vertices_with_correction}
\begin{flalign*}
\nonumber 
\Gamma_{1\sigma \sigma'}^{\alpha \alpha'}  = & \,\, g_{1\sigma \sigma'}^{\alpha \alpha'} + \Big[
2 \, \big( g_{1\sigma \sigma'}^{\alpha \alpha'} g_{2\sigma \sigma'}^{\alpha \alpha'} + \tilde{g}_{1\sigma \sigma'}^{\alpha \alpha'} \tilde{g}_{2\sigma \sigma'}^{\alpha \alpha'} \big)\gamma_{\alpha,\alpha'} \\ 
\nonumber
 & - \big( g_{2\sigma \sigma}^{\alpha \alpha} g_{1\sigma \sigma'}^{\alpha \alpha'} + g_{1\sigma \sigma'}^{\alpha \alpha} \tilde{g}_{1\sigma' \sigma'}^{\alpha \alpha'} \big)\gamma_{\alpha,\alpha}  \\ 
 \nonumber
& - \big( \tilde{g}_{1\sigma \sigma}^{\alpha \alpha'} g_{1\sigma \sigma'}^{\alpha' \alpha'} + g_{1\sigma \sigma'}^{\alpha \alpha'} g_{2\sigma' \sigma'}^{\alpha' \alpha'} \big)\gamma_{\alpha',\alpha'} \\ 
\label{eq:beta-a}  
& + \sum_{\hat{\sigma},\hat{\alpha}} 
 g_{1\sigma \hat{\sigma}}^{\alpha \hat{\alpha}} g_{1\hat{\sigma} \sigma'}^{\hat{\alpha} \alpha'}\gamma_{,\hat{\alpha},\hat{\alpha}} \Big] \left( \ln\left|\frac{\omega}{E_0}\right| - \mathrm{i} \frac{\pi}{2} \right) ,\\ \nonumber
 \Gamma_{2\sigma \sigma'}^{\alpha \alpha'}  = & \,\,  g_{2\sigma \sigma'}^{\alpha \alpha'} + \\ & \big( g_{1\sigma \sigma'}^{\alpha \alpha'} g_{1\sigma \sigma'}^{\alpha \alpha'}  + \tilde{g}_{2\sigma \sigma'}^{\alpha \alpha'} \tilde{g}_{2\sigma \sigma'}^{\alpha \alpha'} \big)  \gamma_{\alpha,\alpha'} 
 \left( \ln\left|\frac{\omega}{E_0}\right| - \mathrm{i} \frac{\pi}{2} \right) ,   \\
 \tilde{\Gamma}_{1\sigma \sigma'}^{\alpha \alpha'}  = & \,\,  \tilde{g}_{1\sigma \sigma'}^{\alpha \alpha'} + 2\, g_{1\sigma \sigma'}^{\alpha \alpha'}  \tilde{g}_{2\sigma \sigma'}^{\alpha \alpha'}\gamma_{\alpha,\alpha'} \left( \ln\left|\frac{\omega}{E_0}\right| - \mathrm{i} \frac{\pi}{2} \right)  , \\
\nonumber 
\tilde{\Gamma}_{2\sigma \sigma'}^{\alpha \alpha'}  = & \,\,   \tilde{g}_{2\sigma \sigma'}^{\alpha \alpha'} + \big(  2\, g_{1\sigma \sigma'}^{\alpha \alpha'} \tilde{g}_{1\sigma \sigma'}^{\alpha \alpha'} + 2 \, \tilde{g}_{2\sigma \sigma'}^{\alpha \alpha'}  g_{2\sigma \sigma'}^{\alpha \alpha'} \\ & 
 -  \tilde{g}_{2\sigma \sigma'}^{\alpha \alpha'}  g_{2\sigma \sigma}^{\alpha \alpha'}  -  \tilde{g}_{2\sigma \sigma'}^{\alpha \alpha'}  g_{2\sigma' \sigma'}^{\alpha \alpha'} \big)\gamma_{\alpha,\alpha'}\left( \ln\left|\frac{\omega}{E_0}\right| - \mathrm{i} \frac{\pi}{2} \right),
\end{flalign*}
\end{subequations}
that leads to the $\beta$-functions of Eq.~\eqref{eq:betafunctions}. Here we introduced the short hand notation $\gamma_{\alpha,\alpha'} = 1/\pi(v_\alpha + v_{\alpha'})$ .

\section{Diagonalizaition of the Gaussian part in the spin space}
\label{sec:app-a}

In order to diagonalize the Gaussian part of the Hamiltonian in the spin space we need to make an orthogonal transformation in the space of the fields $\phi_\sigma$ and $\theta_\sigma$, respectively, where we droped the orbital index for simplicity. The new basis can be constructed with the help of certain generators of the SU(N). The generators of the SU(N) algebra in the fundamental N dimensional representation can be expressed with the help of its peculiar subalgebras: its Cartan subalgebra, and the $\binom{\mathrm N}{2}=\mathrm N(\mathrm N-1)/2$ SU(2) subalgebras.   For the diagonalization we need only the Cartan subalgebra, which is an N$-1$ dimensional algebra of the traceless, diagonal, N$\times$N matrices. The $l$th generators of the Cartan subalgebra can be expressed as: 
\begin{equation}
\label{eq:Cartan-gen}
C_{i}^{(l)}= \left\{
\begin{array}{rl}
1 & \text{if} \,\,\, i\leq l, \\ 
-l & \text{if} \,\,\, i=l+1, \\
0 & \text{otherwise}.
\end{array}
\right.
\end{equation}
Here $i=1\ldots \mathrm N$ and $l=1\ldots \mathrm N-1$, and for simplicity we treat the diagonal matrix as a vector $C_{ii}^{(l)} \equiv C_i^{(l)}$. 

Let us consider an arbitrary spin dependent bosonic field $\phi_\sigma$ with $\sigma=1\ldots \mathrm N$. Now, the transformation defined as 
\begin{subequations}
\label{eq:Cartan-trf}
\begin{flalign}
\label{eq:Cartan-0}
\phi_0 = & \frac{1}{\sqrt{\mathrm N}}\sum_\sigma \phi_\sigma, \\
\label{eq:Cartan-l}
\phi_l = & \frac{1}{\sqrt{l(l+1)}}\sum_\sigma C_{\sigma}^{(l)}  \phi_\sigma \hskip 0.3cm \mathrm{with} \hskip 0.3cm l=1\ldots \mathrm N-1
\end{flalign}
\end{subequations}
will diagonalize any Gaussian Hamiltonian that has SU(N) symmetry in the spin space. The combination \eqref{eq:Cartan-0} itself constitutes the complete symmetric (for the exchange of any two spins) subspace of the spin space, therefore the corresponding excitation modes often called charge or density modes. The combinations  \eqref{eq:Cartan-l} are all orthogonal to the symmetric subspace, they form the antisymmetric subspace of the $\phi$ fields, and they can be referred as spin modes.

\section{Multiparticle umklapp processes} 
\label{sec:umklapps}

In case of $p/q$ commensurate fillings the leading order umklapp processes are multiparticle scattering processes between $q$ fermions \cite{giamarchi04a,buchta07a,szirmai08a}.  The corresponding term of the Hamiltonian has a rather simple form in boson language, it contains cosine terms that couples $q$ phase fields in a fully symmetric manner:
\begin{equation}
\label{eq:umklapps}
\sum_{r_1, \ldots , r_q}  \int \mathrm{d} x \cos (\phi_{r_1}(x) + \cdots + \phi_{r_q} (x) )   .
\end{equation}
Here $r$ denotes the contracted index of all internal degrees of freedom, and the summation has to be understood over the all possible configurations that contain $q$ different internal states. Depending on the relative value of N and $q$, the relevant umklapp terms in Eq.~\eqref{eq:umklapps} mix the various modes. If the total number of the internal states is N, there is no processes with $q>\,$N, because of the Pauli principle. These types of processes are forbidden. If $q=\,$N, there is only one cosine term that contains only the "charge" mode, i.e. the symmetric combination of all the N fields. Contrary, if $q<\,$N, more cosine terms give contribution that couples the various modes. 

As an example, let us consider the two particle umklapp terms, that can be relevant at half-filling. In case of an SU(3) system there are 3 phase fields $\phi_a$, $\phi_b$, and $\phi_c$, and accordingly the two-particle umklapp terms are:
\begin{multline}
\label{eq:two-particle-umklapp}
 \int \mathrm{d} x  \Big[ \cos \big(\phi_{a}(x) + \phi_{b} (x) \big) \\ + \cos \big(\phi_{b}(x) + \phi_{c} (x)\big) + \cos \big( \phi_{a}(x) + \phi_{c} (x) \big) \Big]  .
\end{multline}
The charge and the two spin modes are defined by Eq.\eqref{eq:Cartan-trf} as 
\begin{subequations}
\begin{flalign}
\phi_0 & =\phi_a + \phi_b + \phi_c  , \\ 
\phi_1 & =(\phi_a - \phi_b)/\sqrt{2},  \\ 
\phi_2 & =(\phi_a + \phi_b - 2 \phi_c)/\sqrt{6}. 
\end{flalign}
\end{subequations}
From the above form it is obvious that the cosine terms in Eq.~\eqref{eq:two-particle-umklapp} couples all the 3 modes.

\bibliography{lowdim,ref}

\end{document}